\documentclass[aps,prl,twocolumn,groupedaddress,showpacs]{revtex4}

\usepackage{graphicx}
\usepackage{amssymb}
\usepackage{epstopdf}

\usepackage{amssymb}

\usepackage{graphicx}

\usepackage{color}

\newcommand{\comment}[1]{}

\newcommand{\micron}{\mu\text{m}}

\begin{document}

\bibliographystyle{apsrev}


\title{ Experimental observations of dynamic clustering in dense active colloidal suspensions}
\title{ Dynamic clustering in active colloidal suspensions with chemical signaling}


\author{I. Theurkauff$^1$, C. Cottin-Bizonne$^{1,\star}$, J. Palacci$^2$, C. Ybert$^1$, L. Bocquet$^1$}
\noaffiliation
\email[]{cecile.cottin-bizonne@univ-lyon1.fr}
\affiliation{$^1$ LPMCN,  Universit\'e Lyon 1 and
CNRS, UMR 5586; Universit\'e de Lyon; F-69622 Villeurbanne, France}
\affiliation{$^2$ CSMR, New York University,
4 Washington Place,  New York 10003, USA}
\noaffiliation

\date{\today}

\begin{abstract}
In this paper, we explore experimentally the phase behavior of a dense active suspension of self-propelled colloids. 
In addition to a solid-like and a gas-like phase observed for high and low densities, a novel cluster phase is reported 
at intermediate densities. This takes the form of a stationary assembly of dense aggregates, 
with an average size which grows with activity as a linear function of the self-propelling velocity.
While different possible scenarii can be considered to account for these observations -- such as a generic velocity weakening instability
recently put forward --, we show that the experimental results are reproduced by a chemotactic aggregation mechanism, originally 
introduced to account for bacterial aggregation, and accounting here for diffusiophoretic chemical interaction between colloidal swimmers.
 \end{abstract}


\maketitle


Active systems refer generically to collections of particules 
which consume energy at the individual scale in order to provide self-propelled motion. 
In assembly these systems usually exhibit a wide variety of collective behaviors, structures and patterns, which depart strongly from the classical equilibrium expectations
\cite{Vicsek1995, Llopis2006, Simha2002, chate, Toner2005, Baskaran2009,Tailleur, diLeonardo,Stark, Lowen}. In particular, there are many observations in nature of cluster-like phases in very different systems: flocks, schools, swarms of fishes, insects or bacteria \cite{Budrene1995, Silberzan2011, Dombrowski2004, Joanny2010, Theraulaz2003, Schaller2010}. However this apparent analogy, occuring over a broad range of scales, hides many different mechanisms in the propulsion and interactions.  In order to disentangle the universal from the specific behaviors of those complex phases, a systematic  experimental  exploration of artificial active particles systems at high density is needed. Some experiments have been carried out on self-propelled walkers at high densities \cite{Deseigne2010} but to our knowledge, active particles at the colloidal scale -- involving natural brownian noise and solvent induced interactions -- were only studied at low densities\cite{Golestanian2007, Volpe2011, Palacci2010, mino2011}. 

In this paper we explore experimentally the behavior of a two dimensional dense active suspension of artificial self-propelled colloids (SPC). 
We first characterize the phase behavior of this active system under an external (gravity) field, from the dilute gas to a dense solid-like phase. A key observation is the emergence of dynamic clustering at intermediate densities. Clusters of SPC form naturally in the system, with an average size which grows with activity, in direct proportionality to the propelling velocity of an individual SPC. 
Several scenarii are discussed in order to rationalize these experimental results, suggesting in particular a possible chemotactic aggregation mechanism. 


  

\begin{figure}
\includegraphics[width=7cm,height=7cm]{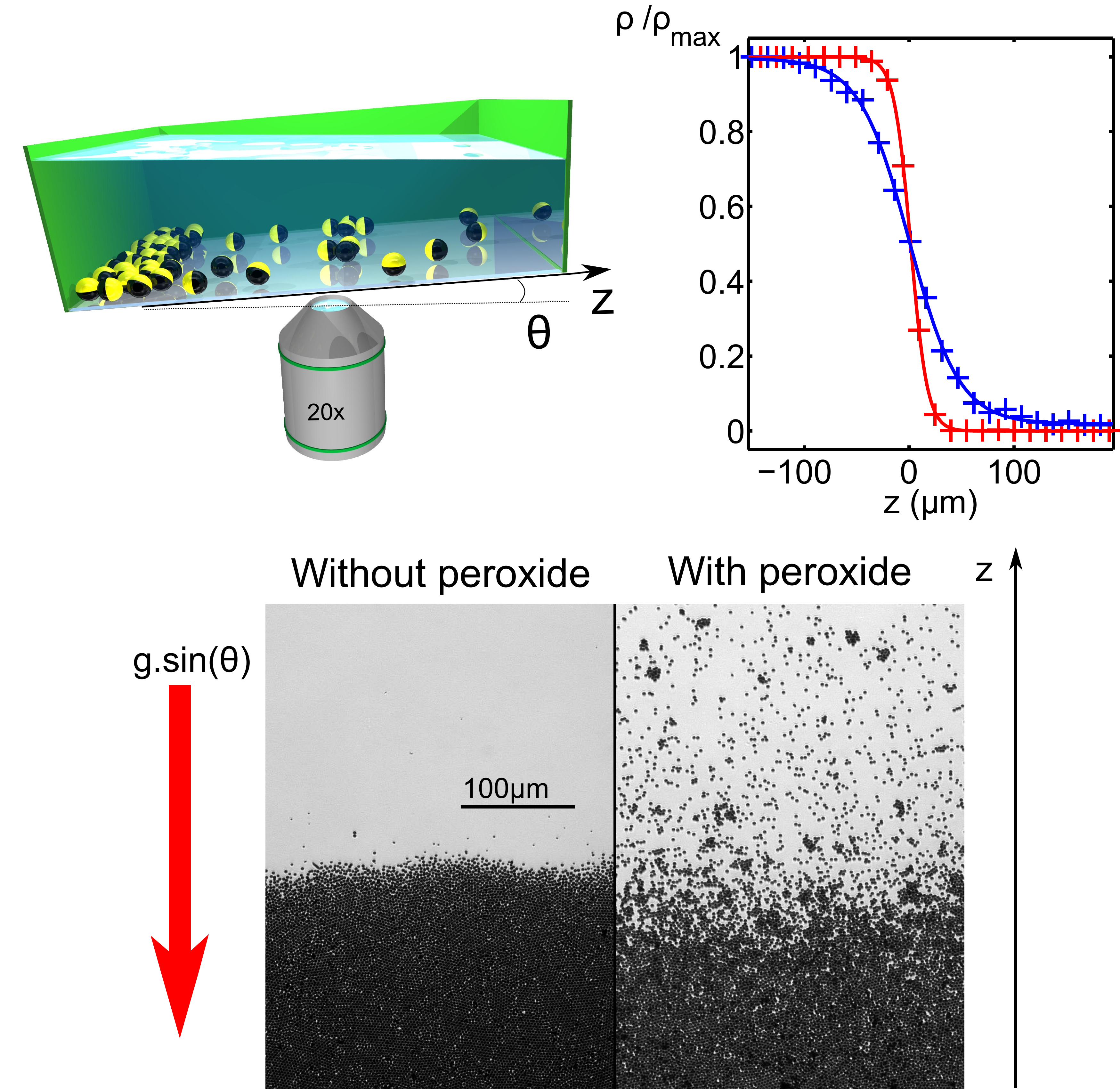}
\caption{Top: (left) Schematic representation of the experimental system; (right) normalized surface density profiles $\rho/\rho_{max}$ 
as a function of the position $\textrm{z}$ along the cell  (with (blue) and without (red)  H$_2$O$_2$). Solid lines are fits with a tangent hyperbolic function. Bottom: picture of the 2D sedimented particles, left: passive colloids in water; right:  active colloids in a solution of  0.1$\%$ of $\textrm{H}_2\textrm{O}_2$.  The effective gravity, $g.\sin \theta \simeq $ 2.$10^{-2}$m.s$^{-2}$ and is indicated by the arrow. The pictures were taken with a 20$\times$ objective.
 }
 \label{fig-1}
\end{figure}

{\it Experimental description --}
The active particles are home-made spherical gold colloids of radius $\textrm{a}\simeq 1\mu$m half covered with platinum \cite{Goia1999}. In presence of hydrogen peroxide the particles self-propel consuming $\textrm{H}_2\textrm{O}_2$ under a self-phoretic motion (a combination of diffusiophoresis and self-electrophoresis \cite{Palacci2010,Posner2011}).  
We note that one key aspect here is that the system does self-propel -- with a typical speed $\textrm{V}\sim3\mu\textrm{m}.\textrm{s}^{-1}$-- even with very low concentrations of hydrogen peroxide. Typical volumetric concentrations of H$_2$O$_2$ are in the range 0.01--0.1\% -- avoiding the formation of $\textrm{O}_2$ bubbles even at high particles concentrations, which is a key issue for alternative active colloidal systems of same family--. The Peclet number for those active particles is typically ${Pe}={{V}.{a}}/{{D}_0}\simeq 13$ where ${D}_0$ is the bare diffusion coefficient of the colloids. 

We observe the system with an inverted optical microscope both in reflection and transmission modes. Videos were taken with a Hammatsu Orca-ER camera at a framerate varying from 1 to 9Hz. 
 A Matlab routine allows particle detection and tracking.  
We confine those particles using  a reduced gravity field: we let the colloids sediment in a cell slightly tilted with an angle $\theta \sim 2.10^{-3}$rad. As the colloids are heavy (with mass density  $\sim19\ \textrm{g.cm}^{-3}$) they quickly sediment at the bottom of the cell and form a dense {\it two dimensional} layer (see Fig.~\ref{fig-1}). As soon as we introduce hydrogen peroxide the particles become active and self-propel, leading to strong change in the sedimentation profile. 
Profiles relax after a few minutes, allowing then to observe and characterize the system in a stable and stationary state over the whole duration of the experiment -- typically a few hours--. 

{\it Sedimentation profile and phase behavior --} 
As observed in Fig.~\ref{fig-1}, the sedimentation behavior of SPC is strongly affected by the amount of activity. 
Non-active SPC particles suspended in water exhibit as expected a solid phase at the bottom, above which a 
very low density gas phase (at room temperature ${T}_{\rm amb}$) can be distinguished, see Fig.~\ref{fig-1}-c (left). When SPC are made active by adding 0.1\% H$_2$O$_2$, Fig.~\ref{fig-1}-c (right), a solid phase remains, but the top gas phase spreads to much higher 
heights, with an exponential decay of the density profile which can be accounted for by a high effective temperature $T_{\rm eff}\sim 50{T}$,  in agreement with the estimate based on the Peclet number, see \cite{Palacci2010}. Furthermore, as can be observed in the experimental snapshots in Fig.~\ref{fig-1}-c, particles tend to cluster for intermediate densities, a behavior which we will exhaustively explore below.

The difference between passive and active particles can also be clearly seen on the surface density profile $\rho$  of the colloids, Fig.~\ref{fig-1}-b. These density profiles are obtained after a binarization of the 500 movie frames (taken at 1fps) 
and averaged over the horizontal direction. The experimental profiles exhibit an enlargement of the ``intermediate zone" between the solid and gas phases when the particles are made active. We finally note that the situation is {\it reversible} which means that when we replace the hydrogen peroxide solution by water, this intermediate zone disappears to recover the non-active sedimentation profile.

\begin{figure}
\includegraphics[width=8cm]{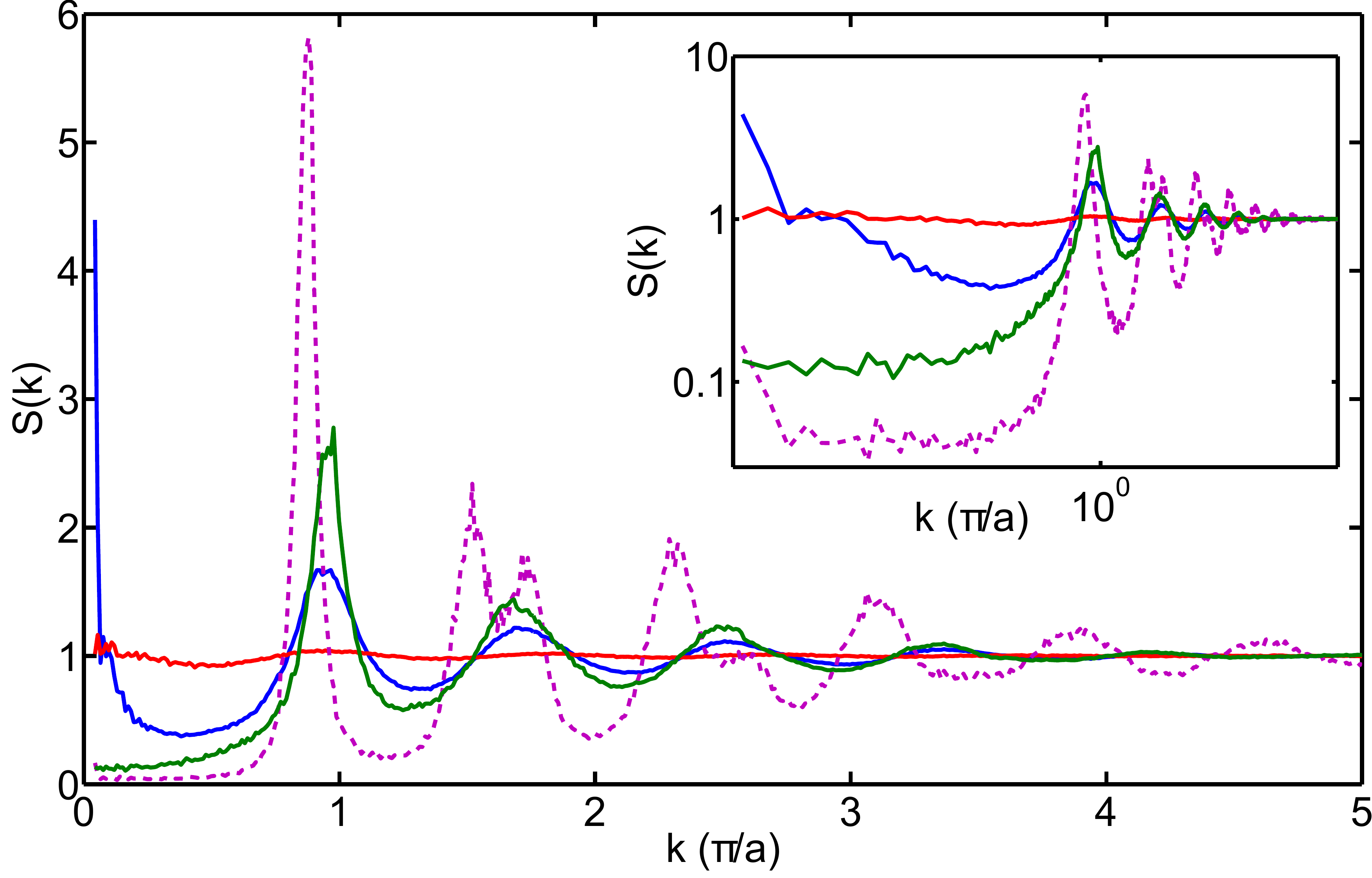}
\caption{Structure factors ${S(k)}$:  Non-active SPC in water in the solid phase, dotted line; Active SPC (in H$_2$O$_2=0.1\%$)  for: solid phase  (green), intermediate zone (blue), gas phase (red). Inset: same curve in logarithmic scales, zooming in the small $k$ behavior.}
 \label{fig-2}
\end{figure}

We then measure the structure factors of the system for various zones in the sedimentation profile, Fig.~\ref{fig-2}.
The structure factor is defined as ${S(k)}=\frac{1}{N}\langle \rho_k \rho_{-k}\rangle $  where $N$ is the number of particles in the region of interest and $\rho_k$ is the Fourier transform of  the instantaneous number density of SPC.  For the SPC without activity -- in water --,  
we report $S(k)$ in the bottom solid phase only. As shown in Fig.~\ref{fig-2} (dotted line), it exhibits a strong ordering characteristic
of a solid phase. It is however not perfectly crystalline due to the size polydispersity of the colloids (typically around 10 $\%$).   For the active SPC, we have measured $S(k)$ in three different zones arbitrarily defined as:  (i) a solid phase -- for  $\rho/\rho_{max}>0.8$ --, (ii) an ``intermediate zone"  -- for $0.05<\rho/\rho_{max}<0.8$ -- and (iii) a gas phase  -- for $\rho/\rho_{max}<0.05$ \footnote{One may note that this threshold value of $\rho/\rho_{max}=0.05$ for the transition between the gas and the intermediate density region is consistent with the value of the density at which one observes a departure of the density profile from the ideal active gas prediction, with a bare exponential decays, see \cite{Palacci2010}.}; $\rho_{max}$ is the maximum surface density (typically of order of the random close packing value). As seen in Fig.~\ref{fig-2}, while
a high ordering remains in the solid phase of SPC, the amplitude of the first peak decreases strongly and activity is observed to destabilize the crystalline phase, 
 in line with recent simulations \cite{Lowen}. For the low density phase, a signature of ordering in the gas phase is also observed (up to 3 peaks can be identified). 
The intermediate ``cluster'' phase is particularly interesting: it exhibits a strong ordering, but  also shows a strong increase of the structure factors when $k\rightarrow 0$, Fig.~\ref{fig-2}. This is a signature of a strong compressibility of the system, reminding of a critical behavior and large fluctuations.

\begin{figure}
\includegraphics[width=8cm]{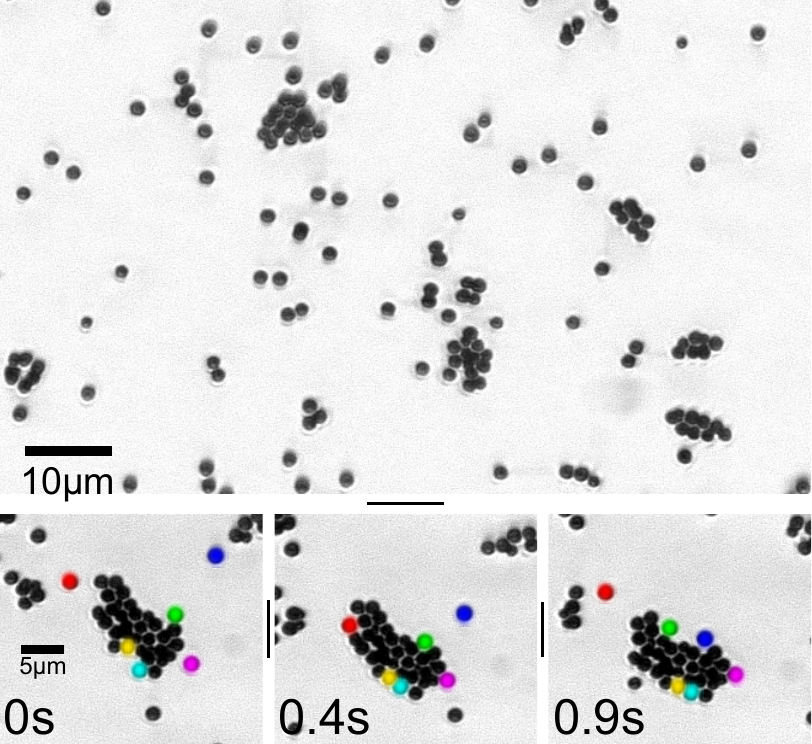}
\caption{Top: Dynamic clusters of SPC observed in a horizontal geometry. The surface fraction of SPC is 5$\%$ and  H$_2$O$_2$ concentration is $0.1\%$. Images obtained in transmission mode with a 63$\times$ objective. Top:  snapshot of the clusters; see \cite{EPAPS}Êfor a movie. Bottom: chronophotography illustrating the dynamics inside one cluster at $t=0$, 0.4 and 0.9 s. Marked particles show a dynamic exchange of SPC between the gas phase and the cluster, as well as internal reorganization of the cluster.}
 \label{fig-3}
\end{figure}

{\it Emergence of dynamic clustering --} 
A key observation in the snapshots, Fig.~\ref{fig-3} and movies \cite{EPAPS}, is the appearance of clustering for active SPC in the regime of intermediate densities, corresponding 
typically to a surface fraction $\phi$ in the range $ 3\%- 50 \%$. 
In order to better characterize this 'cluster phase', we use a non-tilted cell and start with a homogeneous suspension of particles in water, with  surface fraction $\sim$5\%. Upon addition of H$_2$O$_2$,  
we observe after a few seconds the appearance of clusters coexisting with a gas phase, see Fig.~\ref{fig-3}. Those clusters are not immobile but exhibit a slow, random-like, dynamics, with typical speeds of 0.1$\micron.s^{-1}$ to be compared to $ V\sim 3\micron.s^{-1}$ the typical velocity of the SPC. Another striking point is that the clusters are {\it dynamic}, with particles going in and out over time, and clusters merging and dissociating, see Fig~\ref{fig-3} and movieÊ \cite{EPAPS}.   

\begin{figure}[t]
\includegraphics[width=9cm]{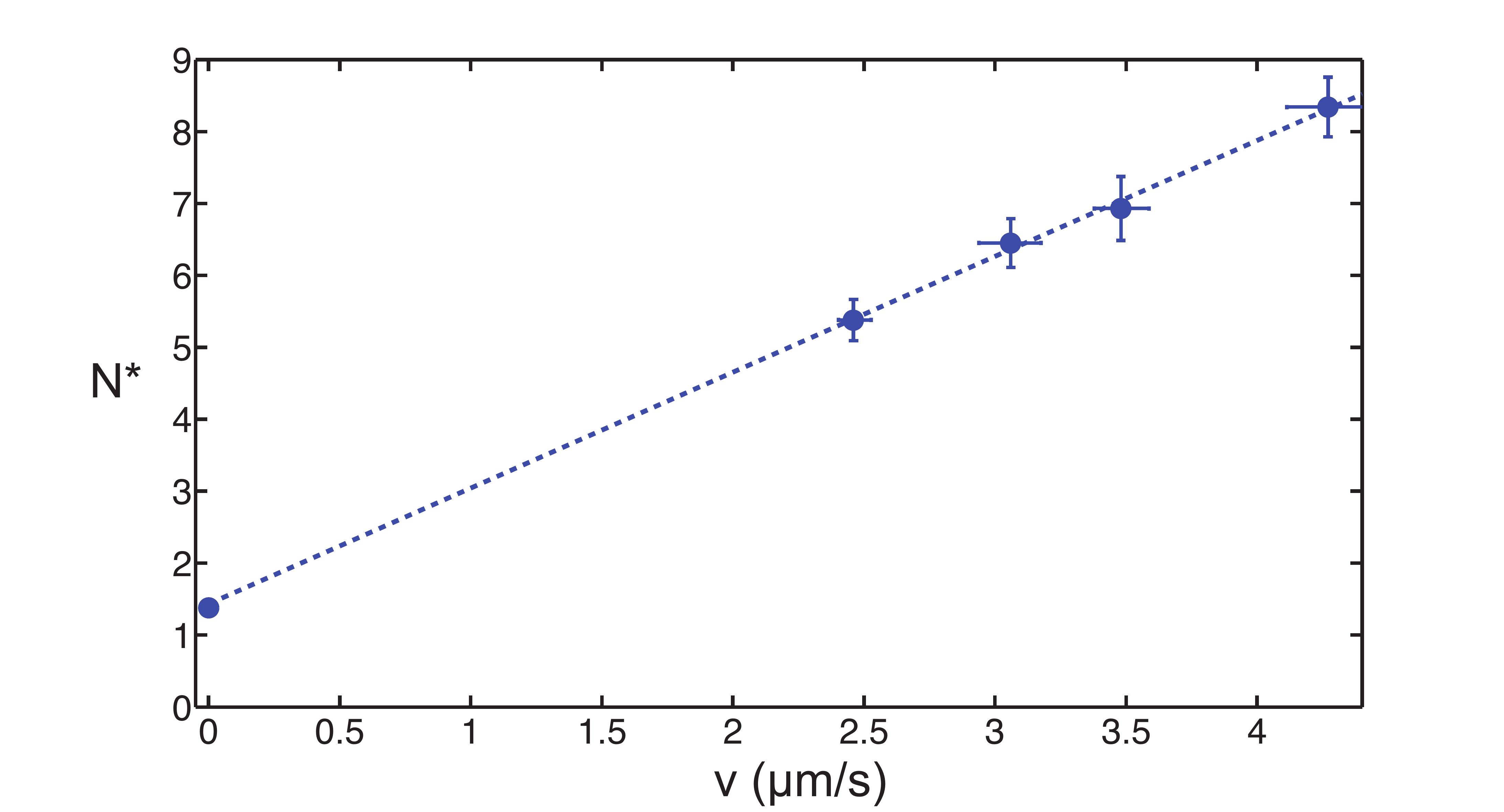}
\caption{Mean cluster size vs  the average velocity of the particles in the gas phase, measured in the experiments (with SPC surface fraction 5\%). The point at $V= 0$ is measured for zero activity. The dashed line is a linear fit,  $N(V) = 1.6\, V + 1.4$
(with $V$ in $\mu$m.s$^{-1}$). }
 \label{fig-4}
\end{figure}

More quantitatively we measured the average size of cluster versus the activity of the system, {\it i.e.} the amount of fuel added, for
volumetric concentration of hydrogen peroxide varying from 0.01\% to 0.1\%. At each H$_2$O$_2$ concentration we measure the activity by tracking the individual SPC in the gas phase and extracting the mean velocity $V$. We count the number of colloids forming each cluster in the stationary state.  A cluster is defined as an assembly of at least three colloids that keeps the same configuration for at least one second. For each hydrogen peroxide concentration, we count more than 150 different clusters. As shown in Fig.~\ref{fig-4}, we measure that the average cluster size $\textrm{N}^*$ is a {\it linear} increasing function of the mean velocity ${V}$ of the individual SPC in the gas phase. This is a counter-intuitive result, as it shows that clustering is {\it enhanced} for larger effective temperature of SPC (as $T_{\rm eff}Ê\sim V^2$ \cite{Palacci2010}).



{\it Tentative scenarii for the dynamic clustering --} 
The theoretical litterature offers various predictions for clustering in active matter, which involve a variety of ingredients and mechanisms \cite{Vicsek1995, Llopis2006, Simha2002, Toner2005, Baskaran2009}. 
We thus try to discriminate between various possible scenarii allowing to rationalize the emergence of clustering in our active SPC, especially for surface fraction as low as a few percents. 
It is accordingly important to first recall some facts about our SPC. 
Our SPC are janus {\it spherical} particles, which leads only to a negligeable anisotropy in the steric interaction between particles, and would {\it a priori} discard modelization of the observed behavior in terms of couplings between density and short-range orientational ordering.
A second important point is that self-propulsion takes its origin in our case in the {\it self-phoretic} motion of the colloids. This leads to a perturbation of the hydrodynamic velocity field around a SPC decaying as $\sim 1/r^3$ \cite{anderson}, in contrast to pusher-puller systems which behave as a force-dipole with a longer  $1/r^2$ decay. As 
 for phoretic motion in general, hydrodynamic interactions are accordingly not expected to play a dominant role in our system.
In contrast, {\it chemical} interactions are susceptible to act as a novel ingredient for our SPC, since, due to the chemical origin of the motion,
the consumption of   fuel and associated production of 'waste' species leads to long range chemical gradient around SPC \cite{Posner2011} -- with concentration fields decaying here like $1/r$ \cite{Golestanian} --. Such gradients are expected to induce diffusiophoretic interactions between particles, as we discuss below.

Among the exhaustive zoology of aggregation behavior discussed in theoretical litterature,
many of the interpretations are difficult to reconcile with the present observations:  \\
- {\it Flocking behavior}, {\it i.e.} group of orientationally ordered particles moving in the same direction  \cite{Vicsek1995, Toner2005}: in our case, SPC clusters do not exhibit directed motion. 
This discards the interpretation in terms of a nematic ordering of the SPC, in line with the expected weak anisotropy of their steric interaction.\\
%
%
%
- {\it Adhesive clustering}: one may {\it a priori}Ê not discard in an obvious way a mechanism based on a simple adhesion between particles, 
by analogy to the structure of simple adhesive colloids \cite{Weitz2006}. However the observed SPC clusters are reversible and strongly kinetic by nature.
This adhesive scenario is also difficult to reconcile with the observation that 
 the size of the SPC clusters {\it increases} with the activity of the particles, Fig.~\ref{fig-4}. \\
%
- {\it Surface generated slip flows}: another mechanism could be the surface-generated attraction between particles induced by  slip flows, as previously described for thermophoresis, Ref.~\cite{Weinert2008,roberto}. 
Such a scenario was shown in  \cite{Weinert2008} to induce two dimensional crystalization of colloids at the surfaces. This is however quite in disagreement with the present observation of kinetically alive clusters, reaching quickly a steady state, stable over hours.

%
While the above scenarii do not provide a convincing framework to understand the present observations, we highlight two alternative views which account better for our experimental results. \\
- First, in a recent numerical study, Fily and Marchetti \cite{Marchetti2012} have shown that self-propelled particles with no alignment mechanism can exhibit an intrinsic clustering instability, though at {\it higher} density. 
They interprete their results in terms of a generic instability taking its origin in the density-weakening dependence of the particles velocities. 
Many features of our experiments are reproduced in their simulations and model, such as the stationary clustering state for intermediate densities and the corresponding strong increase of the structure factor for low $k$, in direct line with 
our Fig.~\ref{fig-2}. 
However at this stage no prediction for the average cluster size $N(V)$ is proposed which could account for the present linear dependence $N(V)\sim V$;\\
- A second possibility can be also drawn, which is directly related to the chemical interactions mentionned above.
The role of chemical sensing has already been studied in the context of patterns and clusters formation of bacteria 
\cite{Keller1970,Budrene1995,Brenner1998,Silberzan2011}. In a pioneering work Keller and Segel (KS) rationalized such behaviors on the basis
of a mean-field description taking into account the diffusion of bacteria, a drift induced by chemical sensing and the production and diffusion of a chemo-attractant \cite{Keller1970,Brenner1998}; its extension to thermotaxic SPC was proposed recently \cite{Golestanian2011}.
%
%
In our context a direct analogy may be made:  the chemotactic effect is played by the {\it diffusiophoretic
motion} \cite{Abecassis} induced by the gradient of species involved in the redox chemical reaction that propels the colloids \cite{Posner2011}. 
We observed independently that 
the consummed H$_2$O$_2$ act as a diffusiophoretic {\it repellent} to colloids, and its reaction induced depletion  is expected to contribute to an overal {\it chemo-attractive} drift between SPC.
Within the KS model, the spatio-temporal dynamics of the active particle population 
can be described at a mean field level in terms of a (here, 2D) particle density $\rho$ and
a global chemoattractant field $c$:
\begin{eqnarray}
&\partial_t \rho &= D_\rho \nabla^2 \rho - \nabla (\mu \rho \nabla c)  \nonumber; \\
&\partial_t c &= D_c \nabla^2 c + \alpha \rho.
\end{eqnarray}
Here $D_\rho$ is the effective diffusion coefficient of the SPC,  $D_c$ is the 'chemoattractant' diffusion coefficient, $\mu$ the diffusiophoretic mobility, $\alpha$ the chemical rate of the powering chemical reaction occuring at the surface of each colloid.


An interesting feature of the KS equations is that they exhibit {\it singular solutions}, leading to a 'chemotactic collapse' of the structure into a single or many dense aggregates (hence termed clumping)  \cite{Brenner1998}. 
This phenomenon introduces a threshold 'Chandrasekhar' number, $N_c$, above which
the bacteria population clusterizes, while below it remains homogeneous. 
The expression for $N_c$ is (in 2D): 
$N_c= {4 D_\rho D_c / \mu \alpha}$.
Furthermore, in the case of clumping -- as discussed in \cite{Brenner1998}  --, $N^\star\approx N_c$ is also expected to fix the typical size of a cluster.
It also interesting to note that $N^\star$ scales like the inverse of the self-trapping coupling constant $g$ introduced by Tsori and de Gennes \cite{deGennes}, so that
for a cluster $g_{\rm eff} = g\cdot N^\star \sim 1$ is at the threshold for self-trapping and reduced mobility.
Here, in order to connect this size to the experimentally measured velocity $V$ of the SPC,
we first note that $V$ is itself a function of these parameters. Indeed  the 
motion is driven by the chemical reaction at the surface of the colloids and following \cite{Golestanian,Posner2011}, 
one expects typically $V \sim \mu \alpha/a D_c$, so that $\mu \alpha \sim V\, a D_c$. Furthermore, the effective
diffusion coefficient of the particles scales like $D_\rho \propto V^2\tau_r$, with $\tau_r$ the rotational brownian
time of the particle \cite{Palacci2010}. Altogether, this yields:
\begin{equation}
N^\star  \sim  {V \tau_r \over a} \sim {Pe},
\label{Nstar}
\end{equation}
where ${Pe}=V.a/D_0$ is a Peclet number characterizing the SPC, which can be also interpreted in terms of 
a persistence number \cite{Taktikos}.
This prediction is in good agreement with the experimental result in Fig.~\ref{fig-3}, $N(V)\propto V$, with furthermore a predicted prefactor given
by an inverse velocity $\tau_r/a \sim \mu$m$^{-1}$.s, also in agreement with the experiments.

This chemotactic scenario, relevant for chemically powered SPC particles, proposes a consistent and predictive explanation for clustering. 
As a coarse-grained and mean-field description, KS cannot describe the detailed kinetics of the aggregate formation and distribution, which remain to be studied
theoretically in more details. It would be also
desirable to explore further the relative
dynamic stability between a collection of cluster versus the relaxation towards a single cluster. Recent simulations of a chemotactic model 
have shown results  similar to the one experimentally observed here with 'hot clusters' coexisting and fluctuating in time \cite{Taktikos}. 
Altogether this framework provides interesting clues to be further explored theoretically in order to rationalize the clustering
mechanism observed in our work.

To conclude, we have explored the high density phase behavior of suspension of active particles at the colloidal scale. 
We have observed the formation of a dynamical ``cluster phase" with a typical cluster size increasing linearly with a Peclet, or persistence, number characterizing the self-propelled colloids. 
Our results can be interpreted in the context
a chemotactic aggregation scenario first introduced to explain clustering observed for bacterial population \cite{Keller1970,Budrene1995}. It suggests that
chemical interactions between SPC can mimic, on a purely physical basis, chemoattractivity and its consequences.
%

We thank B. Abecassis for the synthesis of the gold colloids,  and H. Ayari, M. Kilfoil for the tracking algorithms.
We thank also H. Stark for highlighting discussions.
We acknowledge support from Region Rhone-Alpes under program CIBLE.


\end{document}